\documentclass{article}
\usepackage {amsmath, amsthm, amsfonts, amssymb, amscd, bm, enumerate}
\usepackage[comma]{natbib}
\usepackage{hyperref}
\usepackage{url}
\usepackage[usenames]{color}

\theoremstyle{remark}
\newtheorem{remark}{Remark}
\newtheorem{example}{Example}

\newcommand{\pcite}[1]{\citeauthor{#1}'s (\citeyear{#1})}
\newcommand{\bcite}[1]{\citeauthor{#1}, \citeyear{#1}}

\begin{document}

\title{Switching Cost Models as Hypothesis Tests\thanks{We acknowledge helpful comments of colleagues at the University of
Technology Sydney.}}
\author{Samuel N. Cohen \\ University of Oxford\thanks{Research supported by the Oxford--Man Institute for Quantitative Finance} \and 
Timo Henckel \\ Australian National University \& CAMA \and 
Gordon D. Menzies \\ University of Technology Sydney \& CAMA\thanks{
Gordon D. Menzies (corresponding author), School of Economics and Finance,
University of Technology Sydney, Broadway, Sydney NSW 2007, Australia.
E-mail address: \href{mailto:gordon.menzies@uts.edu.au}{gordon.menzies@uts.edu.au}. Phone: +61-2-95147728. } \and 
Johannes Muhle-Karbe \\ Carnegie Mellon University \and 
Daniel J. Zizzo \\ Newcastle University \& CAMA}

\maketitle

\begin{abstract}
We relate models based on costs of switching beliefs (e.g. due to inattention) to hypothesis tests. Specifically, for an inference problem with a penalty for mistakes and for switching the inferred value, a band of inaction is optimal. We show this band is equivalent to a confidence interval, and therefore to a two-sided hypothesis test.


\end{abstract}

\section{Introduction}

This paper provides a new micro-foundation for two-sided hypothesis tests. Agents receive sequential information and conduct inference which penalizes adjustments to the estimator and deviations from the classical Bayesian estimate. We show that, to a first order approximation for small adjustment costs, the resulting estimator has a band of inaction with width proportional to the Bayesian estimator's standard deviation. This makes it equivalent to a confidence interval and therefore to a two-sided hypothesis test. 

Our result locates belief formation models based on hypothesis tests, such as \pcite{Menzies2009} inferential expectations model, within a wider literature on switching costs due to sticky belief adjustment. Switching costs may arise, say, from  `menu costs', transactions in illiquid markets, cognitive effort in attention and observation or the consultation of experts  (\bcite{Caplin1987}; \bcite{Alvarez2016};  \bcite{Magnani2016}; \bcite{Carroll2003}). 

State-dependent belief adjustments describe how new information about the underlying economic state $X_t$ is incorporated. In our model, agents passively observe until new information exceeds a threshold, depending on the uncertainty of the estimated state, and only then readjust their policy. This infrequent adjustment is similar to models of inattention and portfolio choice (\bcite{Abel2013}; \bcite{Huang2007}).

A concrete example for our analysis is portfolio choice with partial information. Here, $X_t$ represents the unknown expected returns which are estimated from time-series data. If $\hat X_t$ denotes the Bayesian estimate of $X_t$, the optimal portfolio then is typically of the form $h(\hat X_t)$. With transaction costs, this ideal portfolio cannot be implemented and instead has to be replaced by an approximation $h(\Theta_t)$, where $\Theta_t$ is an alternative estimate of $X_t$ that only changes infrequently. The optimal $\Theta_t$ is in turn identified by our tradeoff between switching costs and inefficiency costs due to deviations from the optimal estimator.

Key to our approach is the use of asymptotic approximation methods, to allow closed-form solutions valid when costs are small. This typically yields an approximate `no-action region', within which agents accept deviations from the no-cost optimum  (\bcite{Korn1998}; \bcite{Lo2004}). Our specific contribution is
to link switching cost models to hypothesis tests using the results of \cite{Altarovici2015}. Their purpose is to describe trade within a financial market; we propose that their asymptotic approximation can also be applied to a wide class of recursive estimation problems. 

An appendix contains an illustration of our model in discrete time.

\section{The model}

We base our setting on a Kalman--Bucy filter (\bcite{Kalman1961}), as this has a wide variety of applications (see \bcite{Bain2009}).

We write $X$ for a multivariate (hidden) process, which we seek to estimate using multivariate observations $Y$. We suppose $X$ and $Y$ satisfy
\[
\left\{\begin{split}
dX_t &= F_t X_t dt + dW_t, \qquad X_0\sim N(\hat X_0, P_0),\\ 
dY_t &= A_t X_t dt + dB_t, \qquad Y_0=0,
\end{split}\right.\]
where $W$ and $B$ are independent continuous martingales, with quadratic variations 
\[d\langle W\rangle_t= Q_tdt, \qquad d\langle B\rangle_t = R_tdt.\]
Here $F, A, Q$ and $R$ are matrix-valued deterministic functions of appropriate dimensions, $R$ is invertible and $A$ is nonzero, and $(\hat X_0, P_0)$ are the mean and variance of our initial estimate of $X_0$. The filtration $\mathcal{F}_t = \sigma(Y_{s};s\leq t)$ represents the information available from observing $Y$ up to time $t$.

For these dynamics, conditional on our observations $\{Y_s\}_{0 \leq s \leq t}$, the hidden state $X_t$ has a multivariate normal distribution:\textbf{}
\[X_t|\mathcal{F}_t \sim N(\hat X_t, P_t).\]
The values of $(\hat X_t, P_t)$ have joint dynamics
\begin{equation}\label{eq:filter}
\left\{\begin{split}
 d\hat X_t &= F_t \hat X_t dt + K_t d\hat V_t,\\
dP_t/dt &= F_t P_t + P_t F_t^\top + Q_t - K_t R_t K_t^\top,
\end{split}\right.
\end{equation}
with initial values $(\hat X_0, P_0)$, where $K_t = P_t A_t^\top R_t^{-1}$ denotes the \emph{Kalman gain} process, and $d\hat V_t = dY_t - A_t \hat X_t dt$ defines the \emph{innovations process} $\hat V$, which is a martingale under $\{\mathcal{F}_t\}_{t\geq 0}$, with quadratic variation $d\langle \hat V\rangle_t = R_t dt$. 

\begin{example}\label{example1}
 A simple example is when our processes are all scalar, $F, Q \equiv 0$ and $A, R\equiv 1$. Then $X\equiv X_0$ is a (unknown) constant and $K_t=P_t$, so 
\begin{align*}
 \frac{dP_t}{dt} &= - K_t R_t K_t^\top = -P_t^2 \qquad \Rightarrow \qquad P_t= \frac{1}{1/P_0+t},\\
 d\hat X_t &= K_t d\hat V_t = \frac{1}{1/P_0+t} d\hat V_t.
\end{align*}
Observe that the posterior variance $P_t$ collapses like $1/t$, as we would expect from a standard observation problem. This is the continuous-time analogue of a Bayesian estimation problem for an unknown mean with normal errors, with prior $N(\hat X_0, P_0)$ leading to posterior $N(\hat X_t, P_t)$.
\end{example}

We suppose that, over a fixed time period $[0,T]$, our agent estimates $X_t$ with an approximation $\Theta_t$ of $\hat{X}_t$. She has initial wealth $z$, from which she continuously pays monetary costs $\rho(\hat X_t-\Theta_t)$ due to tracking error relative to the optimal filter estimate and a cost $\lambda$ whenever $\Theta_t$ changes. We assume $\rho$ is convex, smooth and minimized at $\rho(0)=0$. For a utility function $U$, our agent wishes to optimize her utility of expected wealth
\[J(\omega, t, z, \Theta;\lambda) = E\Big[U(Z_T)\Big]= E\Big[U\Big(z-\int_{t}^T \rho\big(\hat X_t-\Theta_t\big) - \lambda \sum_{t\leq s\leq T} I_{\{\Delta \theta_s \neq 0\}}\Big)\Big|\mathcal{F}_t\Big]\]
over piecewise constant adapted processes $\Theta$. As $\hat X$ is a Markov process, there exists a value function
\[v(t, \hat X_t(\omega), z, \Theta_t; \lambda) = \sup_{\Theta':\Theta_t=\Theta_t'} J(\omega, t, z, \Theta'; \lambda).\]

Like in \cite{Korn1998}, \cite{Lo2004} and \cite{Altarovici2015}, the value function can be expanded\footnote{Corresponding verification theorems could be derived using stability results for viscosity solutions as in \cite{Altarovici2015} or martingale methods, cf. \cite{Feodori2016}.} in powers of $\lambda$. If $\lambda$ is small, by ignoring higher order terms, we obtain an approximation to $v$, and hence to the optimal choice of $\Theta$.

\section{Dynamic programming}
With fixed adjustment costs, it  will be optimal to leave $\Theta$ unchanged until $\hat X_t-\Theta_t$ is sufficiently large. Write $\mathfrak{K}$ for the region where $\Theta$ remains fixed. A standard dynamic programming argument yields a partial differential equation for the value function $v(t,\hat x, z, \theta;\lambda)$.

Indeed, the optimal filter $\hat X$ without adjustment costs has the diffusive dynamics \eqref{eq:filter}. By the martingale optimality principle, the value function evaluated along the state variables $(t, \hat X_t, Z_t, \Theta_t)$ is a martingale for the optimal $\Theta$, and a supermartingale otherwise. Applying It\^o's lemma, this implies
\begin{equation}\label{eq:vPDEnoswitch}
 0\geq\partial_t v - (\partial_z v)\rho\big(\hat x-\theta\big) + (\partial_{\hat x} v)^\top F \hat x + \frac{1}{2}\mathrm{Tr}(\Sigma_t\,\partial_{\hat x\hat x} v),
\end{equation}
with equality  on $\mathfrak{K}$ (when it is not optimal to change $\theta$), where 
\begin{equation}\label{eq:sigmadef}
 \Sigma_t = K_t R_t K_t^\top  = P_t A_t^\top R_t^{-1}A_t P_t^\top.
\end{equation}
 Considering the possibility of changing $\theta$, we observe
\begin{equation}\label{eq:vPDEswitch}
 v(t, z, \hat x, \theta;\lambda) \geq \sup_{\theta'} v(t, z-\lambda, \hat x, \theta';\lambda)
\end{equation}
with equality on the complement $\mathfrak{K}^c$ (when it is optimal to change $\theta$). Combining these inequalities, we obtain the dynamic programming equation
\begin{equation}\label{eq:DPE}
 \begin{split}
  0 &= \min\bigg\{-\partial_t v + (\partial_z v)u\big(\hat x-\theta\big) - (\partial_{\hat x} v)^\top F \hat x - \frac{1}{2}\mathrm{Tr}(\Sigma_t\, \partial_{\hat x\hat x} v) , \\
&\qquad \qquad 
v(t, z, \hat x, \theta;\lambda) - \sup_{\theta'} v(t, z-\lambda, \hat x, \theta';\lambda)\bigg\},
 \end{split}
\end{equation}
with terminal value $v(T,z,\hat x, \theta;\lambda) = U(z)$. The difficulty is that the free boundary for $\mathfrak{K}$ needs to be determined as part of the solution. 

\subsection{Asymptotic analysis}
When $\lambda=0$, one can use $\Theta_t=\hat X_t$ to achieve $v(t,z,\hat x; 0) \equiv U(z)$. We expect that the optimal strategy\footnote{This strategy comes from analyzing, over long horizons, how often the boundary of an interval will be hit by a random walk, averaging out the cost paid, then optimizing over the width of the interval chosen. The scaling arguments of \cite{Altarovici2015} and \cite{Lo2004} also apply in our setting, mutatis mutandis. A corresponding analysis in discrete time can be found in the appendix.} will involve switching whenever $|\hat x - \theta|=O(\lambda^{1/4})$, resulting in a cost of $O(\lambda^{1/2})$. This gives the ansatz
\begin{equation}\label{eq:ansatzV}
 v(t, z, \hat x,\theta;\lambda) = U(z)- \lambda^{1/2} \phi(t, z) - \lambda \psi(t,z, \hat x, \xi)+O(\lambda^{3/2})
\end{equation}
where $\xi := \lambda^{-1/4}(\hat x - \theta)$. See \cite{Muhle-Karbe2017} for further discussion.

Recalling our assumptions on $\rho$, 
\[\rho(\hat x - \theta)= \rho(\lambda^{1/4}\xi) = \lambda^{1/2}\xi^\top \Gamma \xi +o(\lambda^{1/2}),\] where
\[\Gamma=\frac{\partial_{xx}\rho(0)}{2}\]
 is a positive-definite matrix. We substitute the ansatz \eqref{eq:ansatzV} into \eqref{eq:vPDEnoswitch}, to obtain
\begin{align*}
0
&\leq\lambda^{1/2} \Big(\partial_t\phi +(\xi^\top \Gamma \xi) U'  +  \frac{1}{2}\mathrm{Tr}\big(\Sigma_t\,\partial_{\xi\xi} \psi\big)\Big) + o(\lambda^{1/2})
\end{align*}
with equality on $\mathfrak{K}$. From \eqref{eq:vPDEswitch} we have
\begin{equation}\label{eq:ansatzswitch}
\begin{split}
0&\leq v(t,z,\hat x, \theta;\lambda) - \sup_{\theta'} v(t,z-\lambda, \hat x, \theta'; \lambda) \\
&=\lambda U'(z)- \lambda\Big(\psi(t,z,\hat x, \xi) - \inf_{\xi'}\psi(t, z-\lambda, \hat x, \xi')\Big) + o(\lambda)
\end{split}
\end{equation}
with equality on $\mathfrak{K}^c$. For small $\lambda$,
\[\inf_{\xi'}\psi(t, z-\lambda, \hat x, \xi')=\psi(t,z-\lambda,\hat x, 0)=o(\lambda).\]
This is because $\psi$ is multiplied by $\lambda$ in \eqref{eq:ansatzV} (so its value at any fixed point is irrelevant to first order) and $\psi$ is smooth. Consequently, \eqref{eq:ansatzswitch} simplifies to
\[0\leq \lambda\Big( U'(z) - \psi(t,z,\hat x, \xi)\Big) + o(\lambda).\]

The leading-order terms for small $\lambda$ in each region in turn lead to the following approximate version of the dynamic programming equation \eqref{eq:DPE}:
\begin{equation}\label{eq:approxDPE}
   0= \min\Big\{\partial_t\phi +(\xi^\top \Gamma \xi) U'(z)  +  \frac{1}{2}\mathrm{Tr}\big(\Sigma_t \,\partial_{\xi\xi} \psi\big),
\quad U'(z) - \psi(t,z,\hat x, \xi)\Big\}.
\end{equation}

\subsection{Exponential Utility}
To obtain a closed-form solution to \eqref{eq:approxDPE}, assume that $U(z)=(1-e^{-kz})/k$ for some $k>0$. Then $U'(z) = e^{-kz}$ and with $\tilde\phi = e^{kz}\phi$, $\tilde\psi= e^{kz}\psi$, \eqref{eq:approxDPE} becomes
\[
   0= \min\Big\{\partial_t\tilde\phi +\xi^\top \Gamma \xi   +  \frac{1}{2}\mathrm{Tr}\big(\Sigma_t \, \partial_{\xi\xi}\tilde\psi\big),
\quad 1 - \tilde\psi(t,z,\hat x, \xi)\Big\}.
\]

Following \cite{Atkinson1995}, we propose\footnote{In one dimension, this is the smallest family of polynomials satisfying our assumptions which are smooth across the boundary.}  a solution of the form 
\[\tilde\psi(t,z,\hat x, \xi) = -1 + (\xi^\top M \xi -1)^2\]
with $\mathfrak{K}=\{\xi: \xi^\top M \xi < 1\}$, where $M$ is a (symmetric, positive-definite) matrix to be determined. We have
\[\partial_{\xi\xi}\tilde\psi = 4(\xi^\top M \xi -1) M - 8 M\xi \xi^\top M.\]
Therefore,
\begin{align*}
 0&= \partial_t\tilde\phi +\xi^\top \Gamma \xi  +  \frac{1}{2}\mathrm{Tr}\big(\Sigma_t \big(4(\xi^\top M \xi -1) M - 8 M\xi \xi^\top M\big)\big)\\
&= \partial_t\tilde\phi -  2\mathrm{Tr}\big(\Sigma_tM\big)+\xi^\top\Big( \Gamma + 2 M \mathrm{Tr}\big(\Sigma_t M\big) - 4  M\Sigma_t M\Big)\xi.
\end{align*}
This has to hold for all $\xi$, so 
\begin{equation}\label{eq:algebraicM}
 0=\Gamma   +  2M\mathrm{Tr}\big(\Sigma_t M\big)- 4 M\Sigma_t M
\end{equation}
which is an algebraic equation for $M$. In one dimension, \eqref{eq:algebraicM} simplifies to $M = \sqrt{\Gamma/(2\Sigma_t)}$,
and using \eqref{eq:sigmadef} the approximately optimal no-switching region is
\[\mathfrak{K}=\Big\{|\xi|\leq \Big(\frac{\Sigma_t}{\Gamma/2}\Big)^{1/4}\Big\}= \Big\{|\hat x-\theta|\leq  \sqrt{P_t}\Big(\frac{A_t}{\sqrt{R_t}}\Big)^{1/2}\Big(\frac{2\lambda}{\Gamma}\Big)^{1/4}\Big\}.\]

\section{Interpretation as hypothesis testing}
We now explore the connection with hypothesis testing in the case where $X$ and $Y$ are scalar processes.

\begin{example}
 Suppose we are in the setting of Example \ref{example1}, so $X_t$ is constant, $F=Q=0$, $A=R=1$, and consider testing the hypothesis
\[\mathrm{H}_0: \mu=\Theta_t \qquad \text{vs.} \qquad \mathrm{H}_1: \mu\neq \Theta_t\]
where $\mu=X_t$ for $t\geq 0$. Recall that
\[P_t = \Sigma_t^{1/2} = (1/P_0 + t)^{-1}\]
is the variance of the hidden process $X_t$ given the observations $\mathcal{F}_t$ until time $t$, and our asymptotically optimal policy is to switch whenever 
\[\frac{|\hat X_t - \Theta_t |}{ \sqrt{P_t}} > c\]
for $c = (2\lambda/\Gamma)^{1/4}$.  By choosing the test size of a two-sided test such that $c$ is the critical value of the usual test statistic, we equivalently switch whenever the standard $z$-test rejects $\mathrm{H}_0$.
\end{example}

In the general scalar setting, the optimal switching region still corresponds to a hypothesis test, but with \emph{variable} test size. Indeed, $\mathfrak{K}$ then has width proportional to
\begin{equation}\label{eq:finalwidth}
 \Sigma_t^{1/4}= \sqrt{P_t} \Big(\frac{A_t}{\sqrt{R_t}}\Big)^{1/2}.
\end{equation}
Observe that $A_t/\sqrt{R_t}$ describes the quality of observations (it is the infinitesimal signal/noise ratio) and hence the volatility of $\hat X$. Therefore, in periods of low-quality data our agent switches more frequently, or equivalently, uses a test with lower confidence level.

\bibliographystyle{abbrvnat}
\bibliography{Switch2}

\section{Appendix: Illustration in discrete time}
In order to give a concrete example of an estimation problem, we consider the task of estimating the parameter $p\in(0,1)$ of independent Bernoulli trials $\{Y_n\}_{n\in \mathbb{N}}$ with values in $\{0,1\}$.  We write $\mathcal{F}_t$ for the information available from the first $t$ observations, that is $Y_1,..., Y_t$. Our analysis of this problem will also yield an asymptotic approximation with the same form as \eqref{eq:ansatzV}. 

The maximum likelihood estimator (MLE) for the parameter $p$ is $\hat{p}_{t}=\sum_{n=1}^t Y_{n}/t$. If $t$ is large, we apply the central limit theorem, and so have the approximate distribution
\[\hat{p}_{t} \stackrel{\scriptsize \mathrm{approx.}}{\sim} N\left( p,\frac{p(1-p)}{t}\right).\]

Consider testing
\[
H_{0} :p=p_{0} \qquad \text{vs.} \qquad H_{1} :p\neq p_{0}.
\]
A two-sided hypothesis test with confidence level $\alpha$ is a rule whereby we maintain 
$H_{0}$ if $\hat{p}_{t}$ falls into a confidence interval. This standard `belief band of inaction' is given by \begin{equation}\label{eq1}
p_{0}-z_{\alpha /2}\sqrt{\frac{p_0(1-p_0)}{t}}<\hat{p}_{t}<p_{0}+z_{\alpha /2}\sqrt{\frac{p_0(1-p_0)}{t}}, 
\end{equation}
where $z_{\alpha/2}$ is the appropriate quantile of a standard normal distribution. We now demonstrate exactly the same $1/\sqrt{t}$ scaling effect from a very different perspective, as in the main body of the paper.

Suppose our agent uses an estimator $\Theta _{t}$, based on the sample $\{Y_{i}\}_{i=1}^t$. She incurs two costs:
\begin{itemize}
\item A cost $\lambda>0$ whenever $\Theta_t$ changes;
\item A cost $\rho(\Theta_t- \hat p_t)$, based on the deviation of $\Theta_t$ from the MLE $\hat p_t$, paid at every time. We assume $\rho$ is twice differentiable, convex and has a minimum at $\rho(0)=0$.
\end{itemize}
We ignore the possible wealth effects and risk aversion, and simply aim to minimize the expected future cost. This is equivalent to $U(z)\equiv z$ (or sending the risk aversion $k\to 0$) in the setting adopted in the main text.

\begin{remark}
The cost $\rho$ can be motivated in various ways. One approach is to treat the true probability $p$ in a Bayesian fashion, and assume our agent faces a running cost $E[(p-\Theta_t)^2|\mathcal{F}_t]$, that is, a cost depending on the distance of their estimate from the true (unknown) value. This is a classic example of a Bayesian loss function, and is suggested by the agent facing additional risk (as measured by the conditional variance) whenever their estimate deviates from $p$.

 In this case, the MLE satisfies $\hat p_t = E[p|\mathcal{F}_t]$, and we can compute
\[E[(p-\Theta_t)^2| \mathcal{F}_t] = E[(p-\hat p_t)^2| \mathcal{F}_t] + (\hat p_t-\Theta_t)^2.\]
As the agent has no control over the term $E[(p-\hat p_t)^2| \mathcal{F}_t]$, the effective cost is given by $(\hat p_t-\Theta_t)^2$, which is of the form considered.

The cost $\lambda$ can be motivated, for example by assuming that the choice of $\Theta$ is an input to a more complex decision setting, which will need to be recalibrated whenever $\Theta$ changes. 
\end{remark}

Related results for models with transaction costs suggest that the optimal policy is for the agent not to act until the tracking error $\hat p_t-\Theta_t$ leaves some interval. To a first-order approximation, which we consider more formally below, the interval is of the form $(-b_t, b_t)$, for some $b_t$ to be determined. When the difference between the MLE and the agent's approximation exceeds this threshold, the optimal strategy is to set $\Theta_t =\hat p_t$ (this is because $\hat p_t$ is an unbiased estimate of $p$ and the adjustment cost does not depend on the size of the adjustment).

To find $b_t$, we will work in an asymptotic regime, where we consider small costs $\lambda\to 0$ and fix a large terminal time $T\to \infty$. In particular, we assume that as $T\to\infty$, there is a time $t_*\to \infty$  such that the long-run costs on the interval $[t_*, T]$ form the principal part of the realized cost,  and costs on the interval $[0,t_*]$ can be ignored.

We first consider the behaviour of the tracking error between two sequential switching times $t_1<t_2$, where $t_*\leq t_1$. (Note that $t_*\to \infty$ as $T\to\infty$, so $t_1\to \infty$ by assumption.) We assume that
\begin{equation}\label{eq:asympassumptions}
 b_{t_1}\to 0,\quad  t_1b_{t_1} \to \infty,  \quad \frac{b_{t}}{b_{t_1}}\to 1\quad \text{and}\quad \frac{tb_{t}}{t_1b_{t_1}} \to 1 \text{ for all }t\in[t_1,t_2]
\end{equation}
 as  $T \to \infty, \lambda\to 0$. We shall see that these assumptions are consistent with the optimal $b_t$ we construct. 

We can write
\begin{equation}\label{eq:randomwalk}
\hat{p}_{t}-\hat{p}_{t_1} = \frac{1}{t}\sum_{i=t_1+1}^{t}(Y_{i}-\hat p_{t_1}) 
\approx \frac{1}{t_1} \sum_{i=t_1+1}^t (Y_i-\hat p_{t_1}),
\end{equation}
where the approximation is justified whenever $t_1^{-1}- t_2^{-1}$ is small, which is justified as $t_1\to \infty$. Assuming $t_1$ is large\footnote{This calculation is for a classical/frequentist approach. With a Bayesian approach we would instead have $P(Y_i=1|\mathcal{F}_{i-1})=\hat p_t \approx \hat p_{t_1}$, giving the same asymptotic approximation.}, we know that 
\[P(Y_i=1)=p \approx \hat p_t \approx \hat p_{t_1},\] so the tracking error is
approximately the sum of a sequence of mean-zero iid random variables, and is well modelled as a random walk, with each step having up-probability $\hat p_{t_1}$ .

For notational simplicity, we write $\hat\sigma_t^2 = \hat p_t(1-\hat p_t)$, which is the estimated variance of our observations. Note that $\hat\sigma_t^2 \to p(1-p)$ as $t\to \infty$, in particular $\hat\sigma_t \approx \hat \sigma_{t_*}$ when $T$ (and hence $t_*$) is large.

Using the approximation of tracking error as a random walk, we choose $b_t$ to minimize expected costs. We have to trade off between our running cost and the cost of switching. For a time $s$, we try and evaluate the expected cost at time $t$, given our barrier strategy $b_t$. We first compute the running cost term.

Write $C_t^\rho(b_t)= \rho(\hat p_t - \Theta_t)$ when $\Theta_t$ is determined using a boundary $b_t$. From our assumptions on $\rho$, provided the tracking error is not too large (which will happen whenever $b_t$ is small), we can approximate with Taylor's theorem $\rho(x) \approx \Gamma x^2$ for some constant $\Gamma>0$.

As $T$ is large,  our agent will be active over a long horizon, so it is the long-run average value of this cost which is important. As $b_t$ will change through time, it is natural to rescale our random walk, so we look for the asymptotic stationary distribution of 
\[\xi_t=(\hat{p}_{t}-\hat{p}_{t_1})/b_{t_1}.\]
This is given approximately by the `triangular' density
\[g(\xi) = \begin{cases} 1+\xi & \text{if } -1<\xi\leq 0,\\ 1-\xi &\text{if } 0<\xi\leq 1,\\ 0 &\text{otherwise,}\end{cases}\]
as can be seen by the observations that:
\begin{itemize}
 \item {$\xi_t$ jumps to zero whenever it hits $\pm b_t/b_{t_1} \to \pm1$, so $g(-1)=g(1) =0$.}
 \item The density integrates to unity.
 \item From considering the possible paths of $\xi$: Except at $x=0$, the only way for $\xi$ to reach $x$ is from being previously at either $x-\frac{1-\hat p_t}{t_1b_{t_1}}$ and observing $Y=1$, or at $x+\frac{\hat p_{t_1}}{t_1b_{t_1}}$ and then observing $Y=0$. The probability of observing $Y=1$ is $\hat p_{t_1}$, so we have the stationary Chapman--Kolmogorov equation
 \[g(x)= \hat p_{t_1} g\Big(x- \frac{1-\hat p_{t_1}}{t_1b_{t_1}}\Big) + (1-\hat p_{t_1}) g\Big(x+\frac{\hat p_{t_1}}{t_1b_{t_1}}\Big)\qquad \text{for }x\neq 0.\] 
 Rearranging and writing $h=1/(t_1b_{t_1})$, we obtain
  \[0= \hat p_{t_1} \frac{g(x- (1-\hat p_{t_1})h)-g(x)}{h^2} + (1-\hat p_{t_1}) \frac{g(x+\hat p_{t_1}h)-g(x)}{h^2}\qquad \text{for }x\neq 0.\] 
Assuming $g$ is twice differentiable for $x\neq 0$ and taking $h\to 0$ (or equivalently $t_1b_{t_1}\to \infty$), this gives the differential equation
  \[\begin{split}
     0&= g'(x) \Big(-\hat p_{t_1}(1-\hat p_{t_1}) + (1-\hat p_{t_1})\hat p_{t_1}\Big) h^{-1}\\&\qquad  + \frac{g''(x)}{2}\Big(\hat p_{t_1}(1-\hat p_{t_1})^2 + (1-\hat p_{t_1})\hat p_{t_1}^2\Big) + o(1)\qquad \text{for }x\neq 0,
    \end{split}
\] 
or equivalently $g''(x) = 0$ for $x\neq 0$. Hence, when $t_1b_{t_1}$ is large (which is assumed when $T\to \infty$, $\lambda\to 0$), we obtain a piecewise linear function for $g$, and hence the triangular density.
\end{itemize}
The density $g(\cdot)$ has variance $1/6$ and substituting, 
\[E[C_t^\rho(b_t)|\mathcal{F}_{t_*}] \approx \Gamma \frac{b^{2}_t}{6}\]
for $t_*\ll t$.

We now seek to understand the expected switching cost,
$E[ C^{\lambda }_t(b_t)|\mathcal{F}_{t_*}]$, where $C^\lambda(b_t) = \lambda$ if  $|\xi_t|\geq 1$ and zero otherwise. { Using our rescaled random walk $\xi$, we need to find the probability of $\xi$ hitting $\pm b_t/b_{t_1} \approx \pm1$ at a time $t\gg t_*$}. As $\xi$ is approximately a random walk restarted at zero, this is approximately $1/E[\tau|\mathcal{F}_{t_1}]$, where $\tau\approx t_2-t_1$ is the time taken to hit $\pm 1$ from zero by $\xi$.

To calculate $E[\tau|\mathcal{F}_{t_1}]$, first observe that $\xi^2_t-ct$ is approximately a martingale for { $c=\hat\sigma_{t_1}^2/(t_1b_{t_1})^2 \approx \hat\sigma_{t_*}^2/(tb_{t})^2$ }and $\xi$ will eventually hit $\pm 1$. By the optional stopping theorem,
\[\xi^2_{t_1}-ct_1 \approx E[\xi^2_{t_2}-{ct_2}|\mathcal{F}_{t_1}] = E[1 -{ct_2}|\mathcal{F}_{t_1}].\]
Rearrangement, and the fact $\xi_{t_1}=0$, yields $1\approx  c E[t_2-t_1|\mathcal{F}_{t_1}] = c E[\tau|\mathcal{F}_{t_1}]$.

For $t_*\ll t$, with $t_1<t\leq t_2$, the probability of hitting the barrier at $t$, thus incurring cost $\lambda $, is { $1/E[\tau|\mathcal{F}_{t_*}]$}, so 
\[E[C^{\lambda}_t(b_t)|\mathcal{F}_{t_*}]\approx \lambda \Big(\frac{\hat \sigma_{t_*}} {tb_t}\Big)^2\qquad \text{for } t_*\ll t.\]

We can now minimize our expectations of long-run future costs. At the leading order, using our approximations, the costs are given by 
\[E[C^{\lambda}_t(b_t)|\mathcal{F}_{t_*}] + E[C^\rho_t(b_t)|\mathcal{F}_{t_*}] \approx \lambda \left(\frac{\hat\sigma_{t_*}}{tb_t}\right) ^{2}+\Gamma\frac{b_t^{2}}{6}.\]
Minimizing this expression in a pointwise manner gives
\[
b_t=\Big(\frac{6\lambda}{\Gamma}\Big)^{1/4} \sqrt{\frac{\hat\sigma_{t_*}}{t}} = \chi \frac{\hat\sigma_{t_*}}{\sqrt{t}}\qquad \text{ where }\chi =\Big(\frac{
6\lambda}{\Gamma}\Big)^{1/4}\hat\sigma_{t_*}^{-1/2}.
\]
Therefore, we obtain a bandwidth $b_t$ for the belief band of inaction that is proportional to $1/\sqrt{t}$, giving the same behaviour as a confidence interval.

As in the continuous time case, we also observe that the width of our band  of inaction is $\chi \hat\sigma_{t_*}/\sqrt{t}$, where $\hat\sigma_{t_*}/\sqrt{t}$ is the standard deviation of the classical estimator. The coefficient $\chi$ is determined by the square root of the signal/noise ratio $\sigma_{t_*}^{-1/2}$. (To see that this corresponds to the signal/noise ratio, observe that $\mathrm{Var}(\hat p_t -\hat p_{t-1}) \approx \hat\sigma_{t_*}/t$, while the variance of $\hat p_{t}$ is $\hat \sigma_{t_*}^2/t$. Their ratio, $\hat\sigma_{t_*}^{-1}$, then corresponds to the term $A_t/\sqrt{R_t}$ from the Kalman--Bucy dynamics \eqref{eq:filter}, which appears in the corresponding band of inaction \eqref{eq:finalwidth}.) 

As in the continuous case, we observe that our band of inaction corresponds with a confidence interval, and hence to a two-sided hypothesis test, with test size dependent on the signal/noise ratio. Using this $b$, we also have 
\[E[C^{\lambda}_t(b_t)|\mathcal{F}_{t_*}] + E[C^\rho_t(b_t)|\mathcal{F}_{t_*}] \approx \frac{\hat\sigma_{t_*}}{t}\sqrt{\frac{2\lambda\Gamma}{3}}\]
which agrees with the $\lambda^{1/2}$ scaling of our asymptotic approximation in continuous time \eqref{eq:ansatzV}.

Finally, we can check the consistency of our choice of $b_t$ with our asymptotic assumption \eqref{eq:asympassumptions}. Clearly, we have $b_t\to 0$ and $tb_t\to \infty$ as $t\to \infty$. We also know that the period between consecutive switches $\tau$ has expectation $E[\tau|\mathcal{F}_{t_*}] \approx (tb_t/\hat\sigma_{t_*})^2 = O(\lambda^{1/2} t)$. As $\lambda\to 0$, this shows that, for consecutive switching times $t_1, t_2$, for $t\in [t_1,t_2]$, 
\[\frac{b_t}{b_{t_1}} \approx\frac{(t_1+O(\lambda^{1/2}t_1))^{1/2}}{t_1^{1/2}} \to 1\quad \text{and}\quad \frac{tb_t}{t_1b_{t_1}} \approx\frac{(t_1+O(\lambda^{1/2}t_1))^{3/2}}{t_1^{3/2}} \to 1.\]
Our choice of $b_t$  is thus consistent with our asymptotic assumption \eqref{eq:asympassumptions} as $T\to \infty, \lambda\to 0$. 

\end{document}